\documentclass[twocolumn,showpacs,preprintnumbers,amsmath,amssymb]{revtex4}
\usepackage{tabularx,graphicx}

\usepackage{color}
\usepackage{hyperref}
\hypersetup{
    colorlinks=true,
    linkcolor=blue,
    filecolor=blue,      
    urlcolor=blue,
}

\usepackage{color}

\usepackage{ulem}   

\begin{document}

\newcommand{\beq}{\begin{equation}}
\newcommand{\eeq}{\end{equation}}
\newcommand{\beqn}{\begin{eqnarray}}
\newcommand{\eeqn}{\end{eqnarray}}
\newcommand{\bmath}{\begin{subequations}}
\newcommand{\emath}{\end{subequations}}
\newcommand{\bra}[1]{\langle #1|}
\newcommand{\ket}[1]{|#1\rangle}

\title{Comment on 
``Trapped flux in a small crystal of
CaKFe$_4$As$_4$ at ambient pressure and in a
diamond anvil pressure cell'' by S. L. Bud'ko et al.}

\author{J. E. Hirsch$^{a}$  and F. Marsiglio$^{b}$ }
\address{$^{a}$Department of Physics, University of California, San Diego,
La Jolla, CA 92093-0319\\
$^{b}$Department of Physics, University of Alberta, Edmonton,
Alberta, Canada T6G 2E1}

\begin{abstract} 
In their paper arXiv:2405.08189, Supercond. Sci. Technol. 37 (2024) 065010 \cite{tiny}, Bud'ko et al. present experimental results for trapped magnetic flux for a tiny sample of 
a type II superconductor. The paper aims to provide evidence in support of the interpretation that 
similar measurements performed in samples of hydrogen-rich materials under high pressure by Minkov, Bud'ko  and coauthors \cite{etrappedp} are conclusive
evidence \cite{ensr2024}  for superconductivity in hydrides under pressure.
Here we point out that the new evidence presented by Bud'ko et al. \cite{tiny}  
further supports our
interpretation \cite{hmtrapped,hmtrapped2} that the reported measurements of trapped flux on hydrides under pressure \cite{etrappedp} are evidence that the
samples are $not$ superconducting.
\end{abstract}

\maketitle 
\section{introduction}

\begin{figure} []
 \resizebox{6.5cm}{!}{\includegraphics[width=6cm]{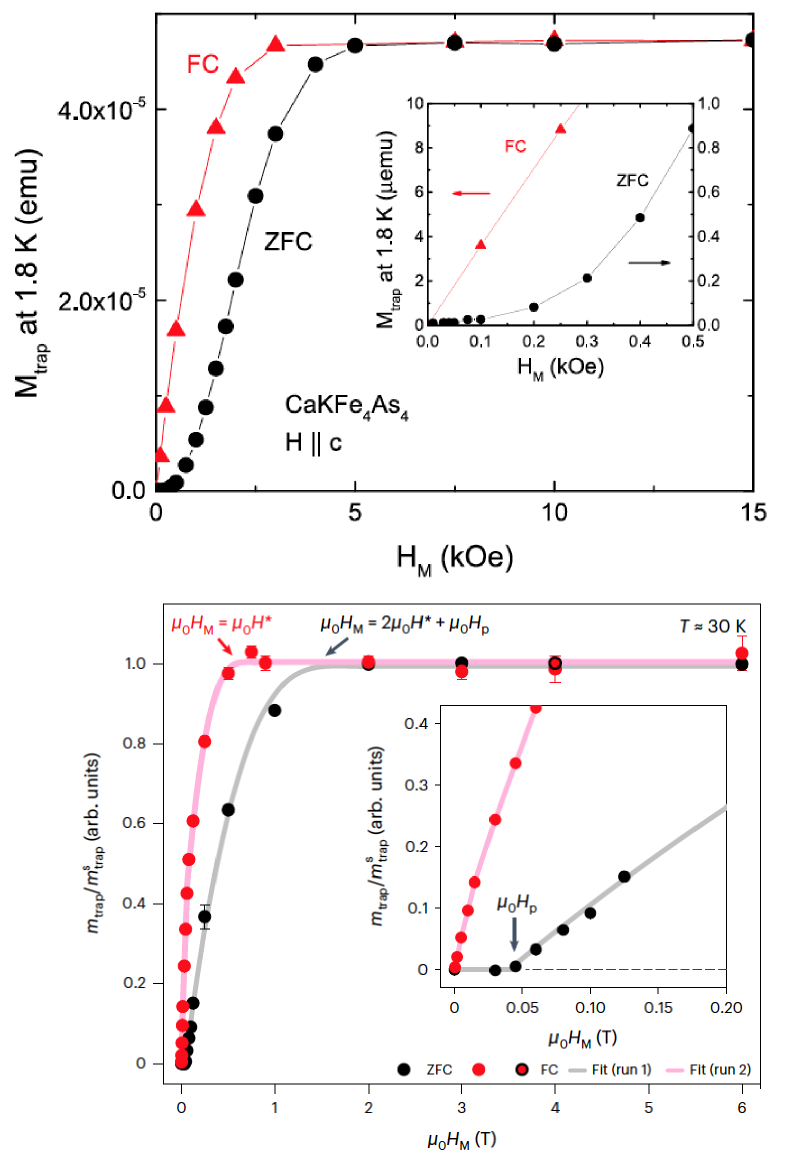}} 
 \caption {Top panel: trapped magnetic moment for 
 a sample of superconducting CaKFe$_4$As$_4$ after applied field is removed in zero field cooling (ZFC, black points) and field cooling (FC, red points)
 protocols, from Ref. \cite{tiny}. Bottom panel: same for a sample of H$_3$S under high pressure, from Ref.~\cite{etrappedp}.
 In both panels, lines were drawn through the points to guide the eye by the authors of Refs. \cite{tiny,etrappedp}. }
 \label{figure1}
 \end{figure}

 In the paper Supercond. Sci. Technol. 37 (2024) 065010 \cite{tiny}, Bud'ko et al. present results of flux trapping measurements
for a tiny sample of superconducting CaKFe$_4$As$_4$. The aim of the work is to provide support for the interpretation that
similar measurements on hydrides under high pressure \cite{etrappedp}  are evidence  that hydrides under high pressure are superconductors.
Figure~\ref{figure1} shows measurements of a tiny sample of the superconducting material on the  top panel, and on the  bottom   panel of a sample of
similar size of H$_3$S. Superficially, the two panels may  look qualitatively similar. In this paper we point out that in fact they are
qualitatively different. But first some historical context.

 Minkov, Bud'ko and coauthors have published results of experiments on hydrides under high pressure \cite{etrappedp}, that showed that when a magnetic field
 is applied to the material and subsequently removed, a remnant magnetization is detected, that they interpret as originating in 
 ``trapped flux''. They inferred from their measurements that the remnant magnetization is produced by electric currents flowing in the 
 material that don't decay with time, providing ``conclusive evidence'' \cite{ensr2024} that the material is a superconductor. 

 In Ref.~\cite{hmtrapped}, we argued that the measurements of Minkov, Bud'ko et al.  \cite{etrappedp} do not support superconductivity because the observed
 dependence of trapped moment under zero field cooling (ZFC) versus magnetic field was linear rather than quadratic. 
 We showed results of our calculations with a simple model showing that the expected behavior due to superconducting currents should
 be quadratic \cite{hmtrapped}. 
 
 Subsequently,  
 Bud'ko, Xu and Canfield \cite{canfield} did similar measurements as Minkov, Bud'ko  et al.  \cite{etrappedp} on known superconducting materials, using samples that were about 1000 times larger in volume
 than the samples used by Minkov, Bud'ko  et al~\cite{canfield}. Bud'ko, Xu and Canfield  concluded \cite{canfield} that their measurements supported the interpretation of the experiments
by Minkov, Bud'ko et al~\cite{etrappedp}.

 Subsequently, Talantsev, Bud'ko et al. \cite{talan} and Bud'ko \cite{budko} criticized our paper Ref. \cite{hmtrapped}, arguing that 
 our paper relied on the ``wrong model'', that it used ``selective manipulations (hide/delete) of calculated datasets'', that we did not perform a proper fit to the Minkov, Bud'ko  et al. measurements  \cite{etrappedp}, that our model was ``unphysical'' and was ``incorrect'', and that its predictions were
 incompatible with the measurements on superconducting samples by Bud'ko, Xu and Canfield \cite{canfield}, hence that our conclusions in
 Ref.~\cite{hmtrapped}   were invalid.
 In these criticisms, they did not address the main point of our paper Ref.~\cite{hmtrapped}, namely that the linear dependence of ZFC moment
 versus field observed in Ref.~\cite{etrappedp} for H$_3$S  is evidence that the signal is not due to superconductivity.
 
Subsequently we showed in Refs.~\cite{hmtrapped2}, \cite{myresponse}  that (i) the criticism by Ref.~\cite{talan} was unfounded, (ii) the criticism by Ref.~\cite{budko} was unfounded,
 and (iii) the reported results by  Bud'ko, Xu and Canfield Ref. \cite{canfield} in fact strongly supported our original analysis and interpretation \cite{hmtrapped} of the
 Minkov, Bud'ko et al. results on hydrides \cite{etrappedp}.
 
 Subsequently, in the paper that we are commenting on here \cite{tiny}, Bud'ko et al. report results on a sample of volume comparable to that of the
 hydride materials, i.e. 1000 times smaller than in their previous work \cite{canfield}, under pressure and at ambient pressure, and argue that the results on their tiny sample
 provide further support to the interpretation that the measurements of Minkov, Bud'ko  et al. on hydrides \cite{etrappedp} are evidence for superconductivity.
 In Ref.~\cite{tiny}, Bud'ko et al also repeat their claim \cite{budko} that  their new results  contradict the ``unphysical'' and ``incorrect'' results
 of Ref.~\cite{hmtrapped}.

  \section{analysis}
  
  Our analysis of flux trapping experiments in Refs.~\cite{hmtrapped,hmtrapped2} was based on the Bean model, under the simplifying 
  assumption that the critical current is uniform over the superconducting region. We recognize that this is a simplifying assumption and
  that real superconducting samples could show some variation of critical current as function of field magnitude and position in the sample,
  so we don't expect exact agreement of calculated results with measured results. Still, we expect that our model will capture the essential physics
  and yield approximately correct results for  measurements of trapped flux in superconducting samples.
  In particular, an important aspect of the physics emphasized in Refs.~\cite{hmtrapped,hmtrapped2} is that under ZFC the trapped
  magnetic moment predicted by our model and mandated on physical grounds should be quadratic and not linear as function of applied field.
  
  The model, quantitatively described in Refs.~\cite{hmtrapped,hmtrapped2}, has the following adjustable parameters:
  (i) $H^*$, the value of the magnetic field for which an applied magnetic field reaches the center of the sample, which is proportional to the
  critical current density. (ii) $H_p$, the threshold value of the magnetic field below which an applied field does not penetrate, which is the
  lower critical field corrected for demagnetization, and 
  (iii) the value of the saturation trapped magnetic moment $m_s$.

 In this Comment, we show that in fact the most recent  results of Bud'ko et al.~\cite{tiny} provide further confirmation of the validity of our analysis \cite{hmtrapped,hmtrapped2} of the 
 Minkov, Bud'ko et al. experiments \cite{etrappedp} and of our interpretation that the signals measured in hydrides are not due to superconductivity.
 
     \begin{figure} []
 \resizebox{8.5cm}{!}{\includegraphics[width=6cm]{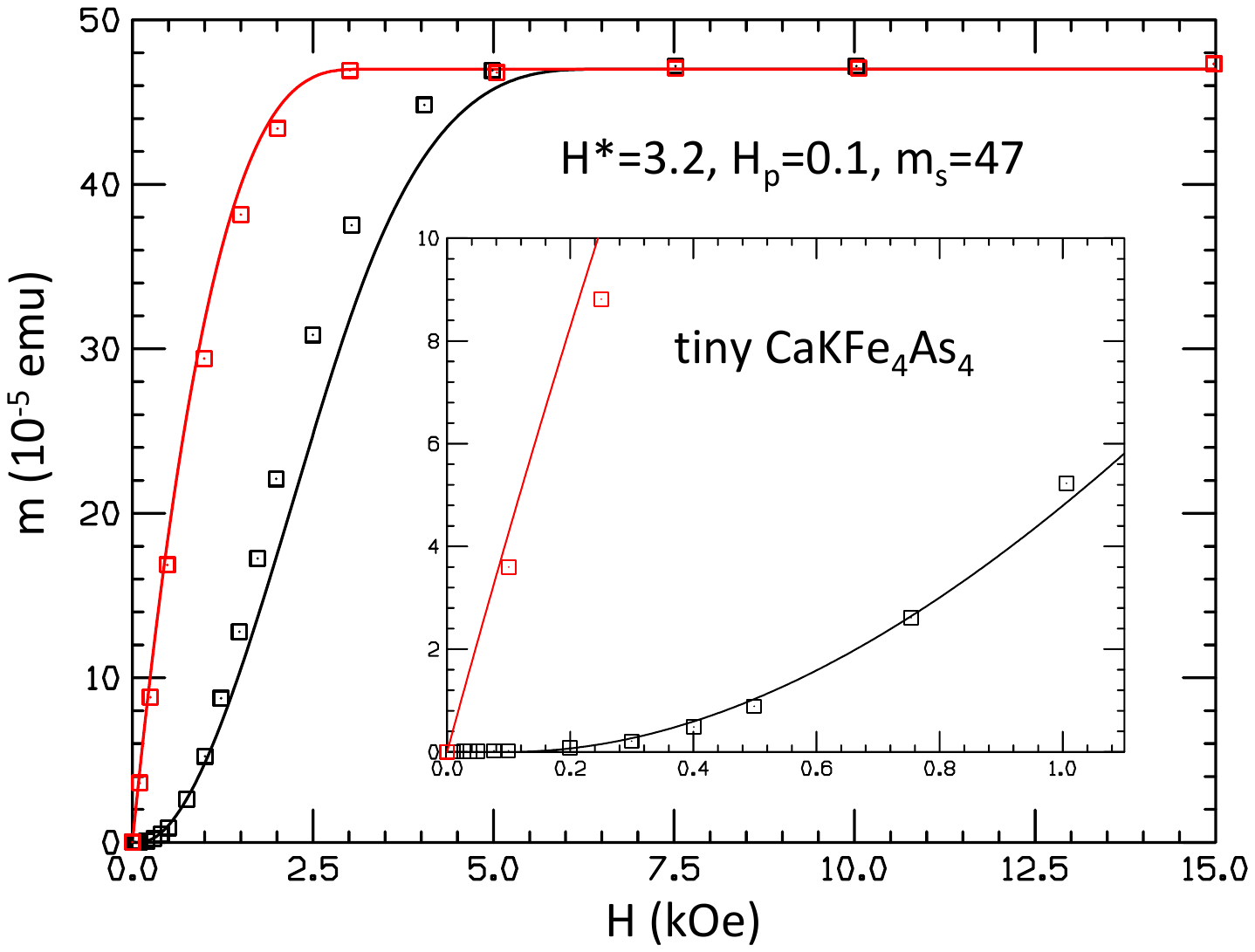}} 
 \caption { Theoretical fit to the experimental results of Ref.~\cite{tiny} for trapped flux in CaKFe$_4$As$_4$ (red and black lines) using
 the theoretical model of Refs.~\cite{hmtrapped,hmtrapped2}. Parameters used for the fit are given in the figure.
}
 \label{figure2}
 \end{figure} 
 
  Figure~\ref{figure2} shows comparison of the predictions of our model with the measured results on the superconducting sample studied in Ref.~\cite{tiny}, with the
  values of the adjustable parameters shown in the figure. These values were chosen so as to give the best possible fit to the 
  low field values of the ZFC trapped moment for low fields, which clearly shows supralinear behavior (quadratic) as seen in the
  inset of Fig.~\ref{figure2}. It can be seen that these parameters also give a good fit to the FC moment at low field (inset of Fig.~\ref{figure2}).
  Furthermore, they provide a reasonable fit to the behavior over the entire field range, as shown in the main body of Fig.~\ref{figure2}.
  There are some  deviations from the measured values at intermediate field values, but it can be seen that the values where the magnetic
  moment reaches saturation, which in our model is $H^*$ and $2H^*+H_p$ for FC and ZFC protocols respectively, are not far from
  the values indicated  by the experimental points.
  
     \begin{figure} []
 \resizebox{8.5cm}{!}{\includegraphics[width=6cm]{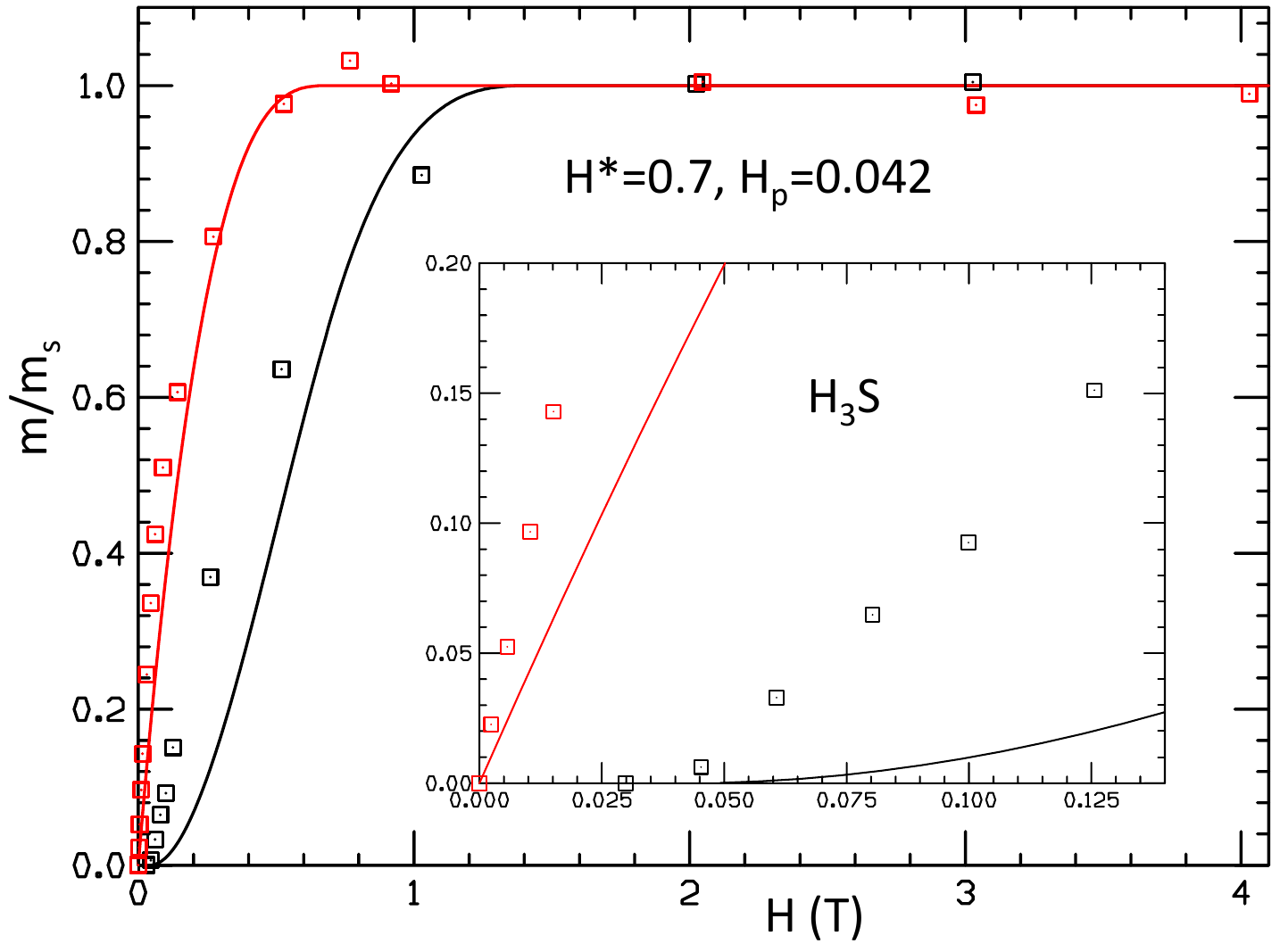}} 
 \caption {  Theoretical fit to the experimental results of Ref.~\cite{etrappedp} for trapped flux in H$_3$S (red and black lines) using
 the theoretical model of Refs.~\cite{hmtrapped,hmtrapped2}. Parameters used for the fit are given in the figure.
 They were chosen to approximately match the field values where the magnetic moment reaches saturation.
}
 \label{figure3}
 \end{figure}

     \begin{figure} []
 \resizebox{8.5cm}{!}{\includegraphics[width=6cm]{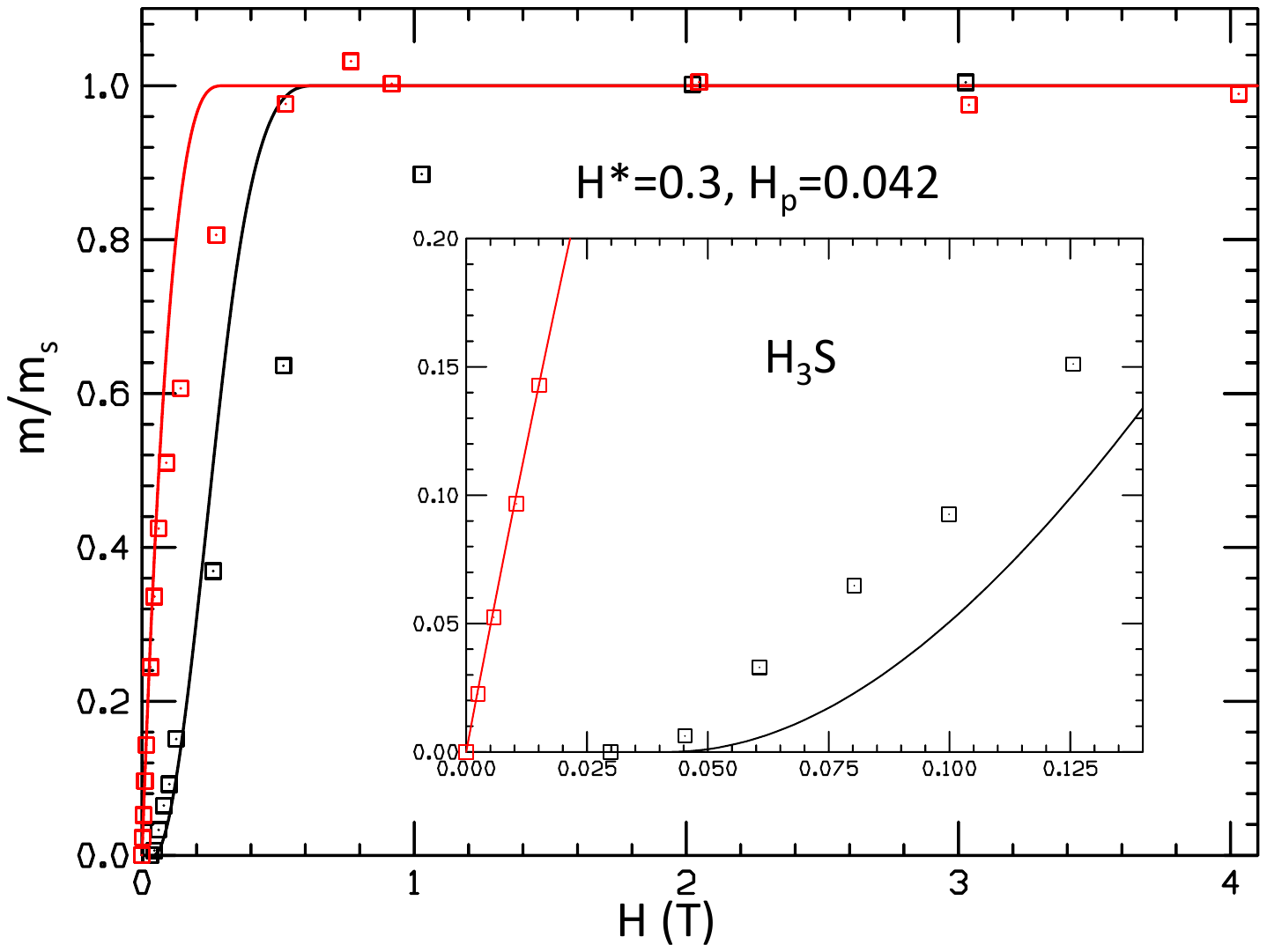}} 
 \caption { Theoretical fit to the experimental results of Ref.~\cite{etrappedp} for trapped flux in H$_3$S (red and black lines) using
 the theoretical model of Refs.~\cite{hmtrapped,hmtrapped2}. Parameters used for the fit are given in the figure.
 They were chosen to approximately match the low field behavior of the FC moment.
}
 \label{figure4}
 \end{figure}

     \begin{figure} []
 \resizebox{8.5cm}{!}{\includegraphics[width=6cm]{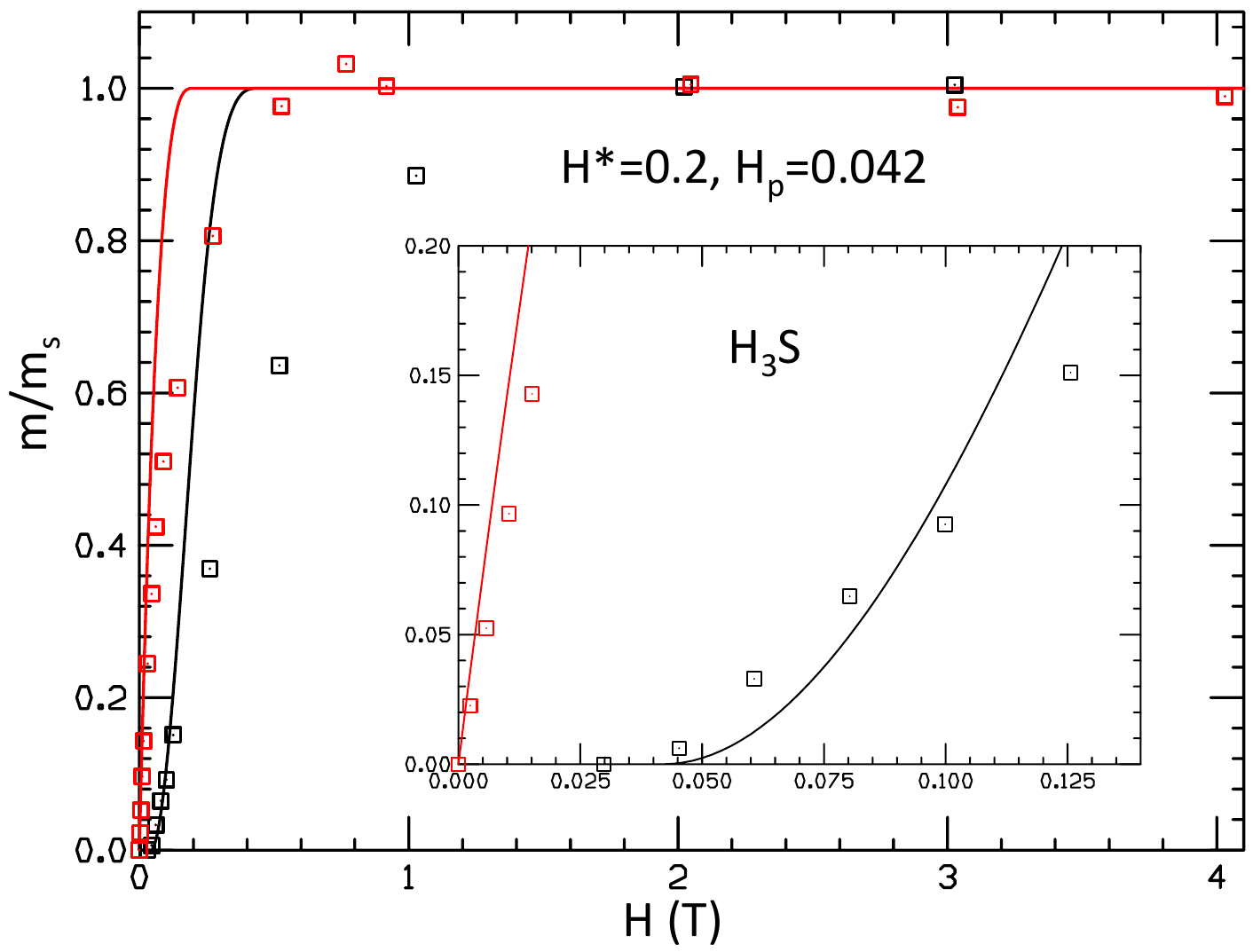}} 
 \caption { Theoretical fit to the experimental results of Ref.~\cite{etrappedp} for trapped flux in H$_3$S (red and black lines) using
 the theoretical model of Refs.~\cite{hmtrapped,hmtrapped2}. Parameters used for the fit are given in the figure.
 They were chosen to approximately match the low field behavior of the ZFC moment.
}
 \label{figure5}
 \end{figure} 
 In Figs.~\ref{figure3},\ref{figure4},\ref{figure5} we attempt such fits for the hydride sample $H_3S$ using the measured values reported in Ref.~\cite{etrappedp}. The low field behavior of the ZFC moment gives a value
 of $H_p=0.042T$, as determined in Ref.~\cite{etrappedp}, which we use in the three figures. We start in Fig.~\ref{figure3} by choosing $H^*$ to 
 approximately fit the field values where the moment reaches saturation, which was determined to be $H^*=0.7T$ in Ref.~\cite{etrappedp}.
  The moment under ZFC reaches saturation at magnetic field approximately twice as large as the FC, as our model predicts. It can be seen in Fig.~\ref{figure3} that the red and black lines approximately fit
 the measured points near the values where the moments reach saturation. However, it can be seen in the inset of Fig.~\ref{figure3} that there
 is drastic disagreement with the behavior of the moments at small field, the theoretical curves giving values 
 substantially lower than the measured ones, particularly for the ZFC protocol.
 
 In Fig.~\ref{figure4} we pick $H^*$ so as to fit the low field behavior of the FC moment. This requires a much smaller value of $H^*$, namely 
 $H^*=0.3T$. It can be seen that the low field values of the ZFC moment are still substantially underestimated. Furthermore, the overall
 behavior seen in the main body of the figure is in drastic disagreement with experiment, since the curves reach saturation
 substantially before the measured moments reach saturation, both for FC and ZFC protocols.
 
 Finally, in Fig.~\ref{figure5} we pick an even smaller value of $H^*$, $H^*=0.2$, to attempt a better match to the low field ZFC moments.
 It can be seen that the ZFC moments are still not matched by the theory, because they don't follow the same functional form as
 the theory predicts, namely quadratic in field, nevertheless we try to fit the behavior on average, with some points falling below, some above,
 the theoretical curve. The small field FC moment is now overestimated by the theoretical values,
 and as seen in the main panel of the figure the overall disagreement with the measured values is stark both for
 FC and ZFC moments.
 
      \begin{figure} []
 \resizebox{8.5cm}{!}{\includegraphics[width=6cm]{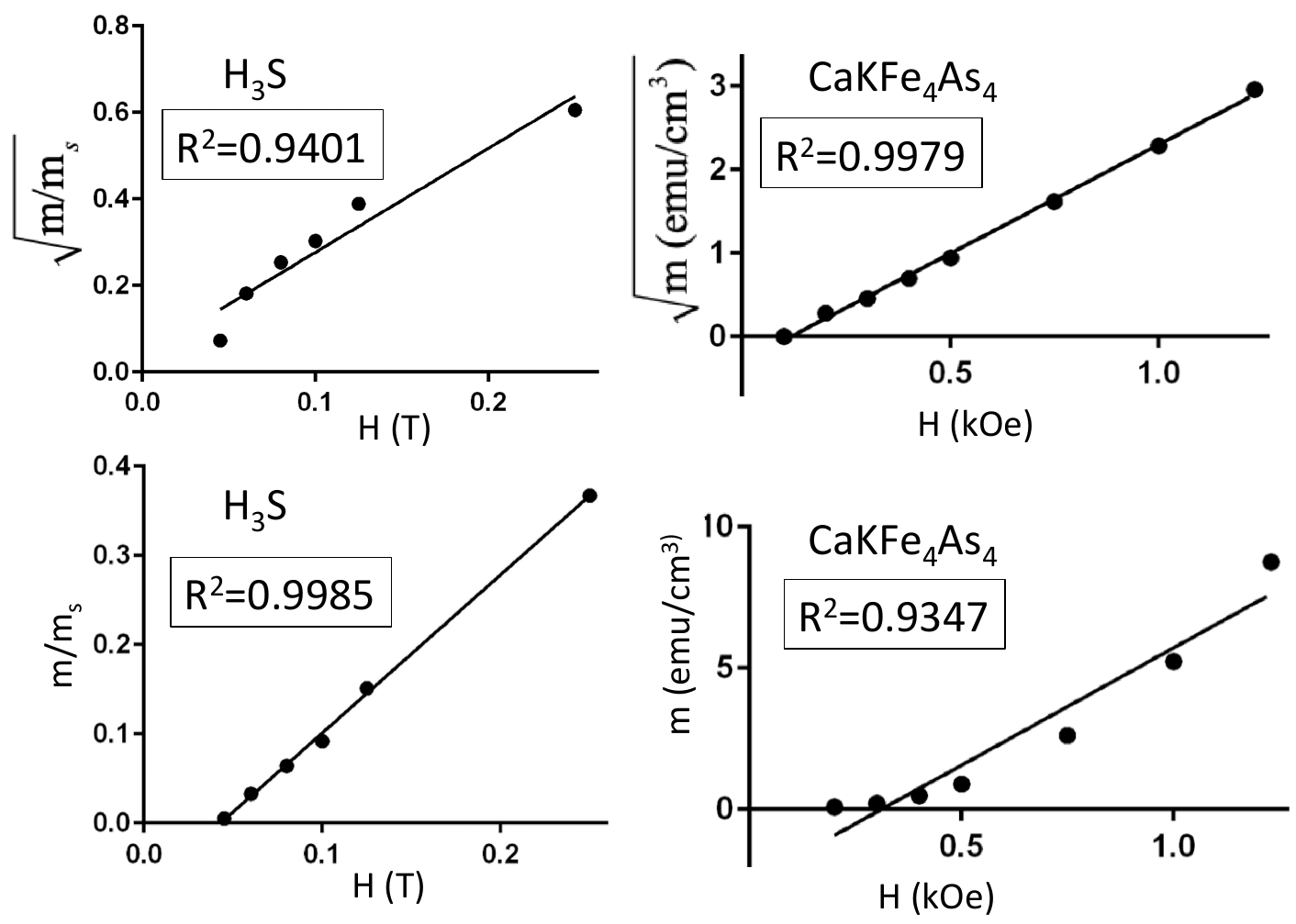}} 
 \caption { Top panels: Linear regression fits of square root of trapped moment versus magnetic field for H$_3$S~\cite{etrappedp}  (top left panel) and
 for the tiny sample of  CaKFe$_4$As$_4$~\cite{tiny} (top right panel). 
 It can be seen that for the  superconductor (top right panel) the fit is very good, indicating quadratic behavior of
 trapped moment versus field as predicted by our model and by general physical arguments. Instead,
 for H$_3$S (top left panel)  the fit is not good.
Bottom panels: Linear regression fit of trapped moment versus magnetic field for H$_3$S (bottom left panel) and
 for CaKFe$_4$As$_4$ (bottom right panel). 
 It can be seen that for the superconductor (bottom right panel)  the fit is not very good, indicating that the trapped moment under
 ZFC is not linear with magnetic field,  as predicted by our model and general physical arguments. Instead,
 for H$_3$S (bottom left panel) the linear  fit is excellent.
}
 \label{figure6}
 \end{figure}

 Finally, as we also did in Ref.~\cite{hmtrapped2},
 in order to quantify the deviation of the low-field ZFC data for H$_3$S from the expected behavior of a superconductor, we show in 
Fig.~\ref{figure6} linear regression fits \cite{regression} to the square root and first power of the trapped moment versus magnetic field
respectively for the two materials  considered here. The coefficient $R^2$ measures how well the data fit the regression model,
with $R^2=1$ being a perfect fit. It can be seen in the top panels that $R^2$ is larger than 0.99 for the superconductor
indicating an excellent fit to quadratic behavior of magnetic moment versus magnetic field, while for 
H$_3$S the $R^2$ differs substantially from unity, indicating the observed behavior is not consistent with quadratic dependence of 
moment on field. Conversely, as seen in the bottom panels, a linear fit of magnetic moment vs field 
fits the H$_3$S data extremely well, with $R^2=0.9979$, and does substantially worse for the standard
superconductor. This provides strong evidence that for the standard superconductor the behavior of magnetic moment
versus field under ZFC is quadratic while  it is linear for H$_3$S, just as we found in Ref.~\cite{hmtrapped2} for several
other  superconductors as well.

\section{effect of pressure and background subtraction}

\begin{figure} []
 \resizebox{8.5cm}{!}{\includegraphics[width=6cm]{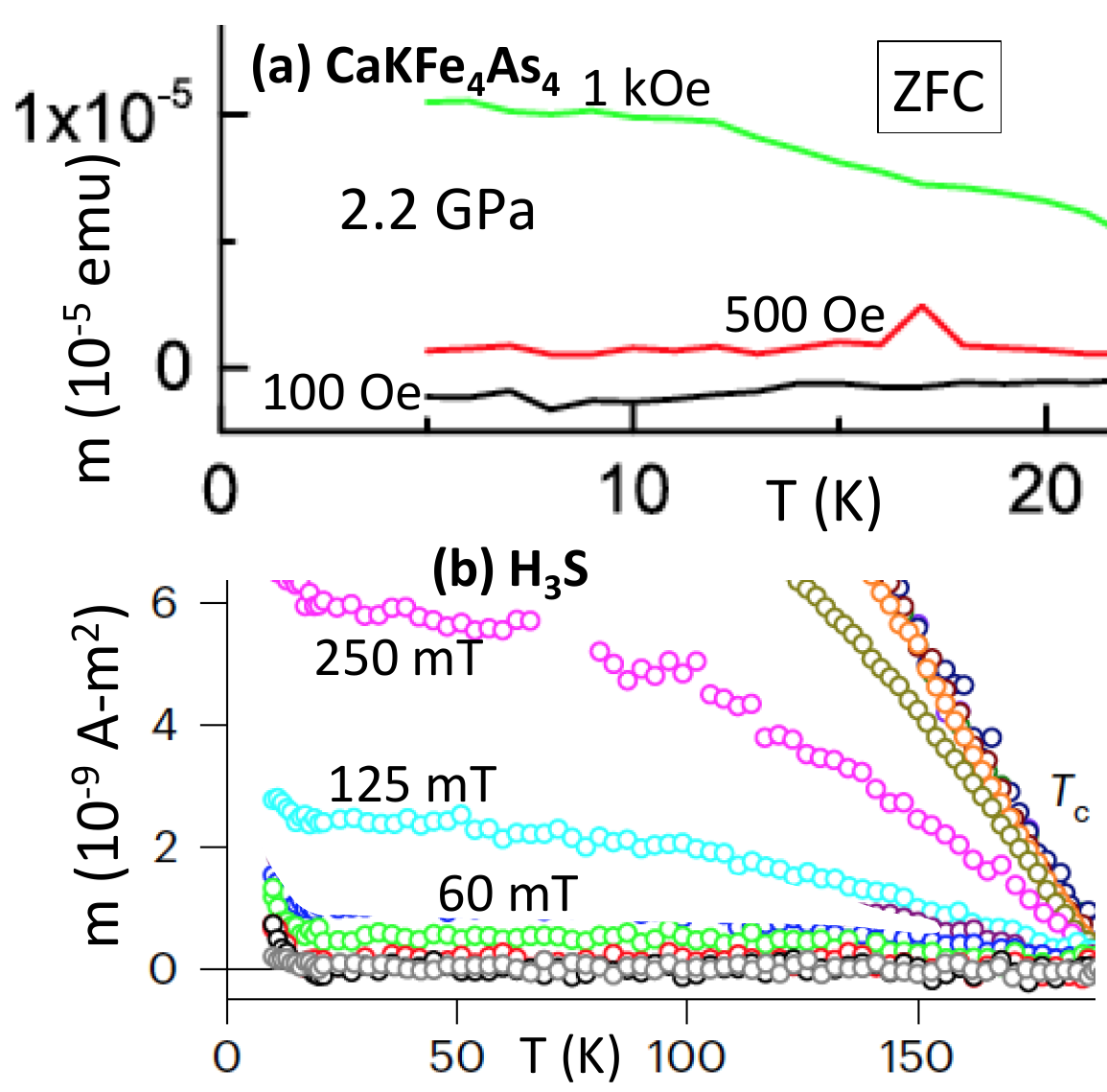}} 
 \caption {  (a) From Fig.~5(b) of Ref.~\cite{etrappedp}: temperature dependent trapped flux magnetization of CaKFe$_4$As$_4$ measured in H =0 using  ZFC protocol in DAC
under pressure of 2.2 GPa. The magnetic field was applied at base temperature $T=1.8K$, then removed.
(b) From Fig.~1(c) of Ref.~\cite{etrappedp}:
 temperature-dependent trapped flux magnetization of H$_3$S measured in ${\rm H} =0$ using  ZFC protocol in DAC
under pressure of 155 GPa. The magnetic field was applied at base temperature $T=10K$, then removed.
}
 \label{figure7}
 \end{figure}

For the sample under pressure, Ref.~\cite{tiny} does not report magnetic field dependence of the trapped moment  with sufficient detail
in their Fig.~6   to make it possible 
for us to fit the results as done in Fig.~\ref{figure2}. Nevertheless, we can get some information from the temperature-dependent  
trapped moments reported in Fig.~5(b) of Ref.~\cite{tiny}, shown here in Fig.~\ref{figure7}(a). It is very clear from Fig.~\ref{figure7}(a)  that the behavior of
the ZFC moment with field in the DAC under pressure  is very non-linear, just as it was for the sample at ambient pressure: in increasing the magnetic
field from 500 Oe to 1 kOe, i.e. by 500 Oe, the increase in trapped moment seen in Fig.~\ref{figure7}(a)  is a factor of 5 times larger  than in increasing the
magnetic field  from 100 to 500 Oe, i.e. by 400 Oe. This is in clear disagreement with what would be expected
for a sample whose trapped moment  behaved linearly like the hydride sample shown in the inset of Fig.~\ref{figure1} lower panel, which would show an 
increase in moment when going from 500 Oe to 1 kOe that is only a factor 1.25 larger than in going from 100 to 500 Oe.
For comparison, Fig.~\ref{figure7}(b) shows similar curves for H$_3$S from Fig.~1(c) of Ref. \cite{etrappedp}. 
Here, the increase in the low temperature moment in going from 125 mT to 250 mT, i.e. an increase of 125 mT,  is approximately 
twice the increase in going from 60 mT to 125 mT, i.e. an increase of 65 mT, as expected for linear behavior.
 
Furthermore, while the purpose of Ref.~\cite{tiny} is ``{\it to address concerns about sample size and signal associated with 
trapped flux measurements in a DAC},'' i.e. under pressure, the authors of Ref.~\cite{tiny} have actually managed to raise 
concerns instead. They themselves have noted the large discrepancy in signal size in the normal state 
under 2.2 GPa pressure in the DAC (their Fig.~4(a))  compared to that
at zero pressure without the DAC (their Fig.~1), and the accompanying increase in noise, possibly attributing these discrepancies to the presence of
``{\it a fairly large background}'' in the DAC (see the discussion on p. 4-5 of Ref.~\cite{tiny}). The authors include an Appendix B 
with figures produced from a commercial DAC --- see the description of this  commercial DAC plus accompanying 
software in Ref.~[16] of Ref.~\cite{tiny}, --- but do not show background and sample + background results as a function of temperature
or applied magnetic field. It would also have been useful to see more detailed results for the trapped field magnetization
as a function of applied field in the DAC under pressure  --- only four unsaturated values are shown in their Fig.~6.

As one of the purposes of Ref.~\cite{tiny} is to convince the reader of the validity of hydride results 
presented in Ref.~\cite{etrappedp} (their Ref.~[6]),
it is unfortunate that there are ``{\it important differences in protocols for trapped flux magnetization measurements between this work and
[6].}'' In particular the presence of a ``{\it remnant magnetic field...which depends on the superconducting magnet design and geometry
as well as the history of magnetic field applied prior to the measurement}'' is a disturbing source of uncertainty. One is left wondering
about the possibility of spurious effects in both these measurements and those in the hydrides, Ref.~\cite{etrappedp}, that could
depend significantly on the background subtraction procedure that is applied, that is not clearly explained in 
Ref.~\cite{etrappedp}.

Given that this group has considerable experience with
CaKFe$_4$As$_4$ at even higher pressures (above 4 GPa) where this material is {\it not} superconducting \cite{kaluarachchi2017}, 
it would be of considerable interest to test for trapped flux under these higher pressures   using the same background subtraction procedure as used for the lower pressures where it is superconducting. 
 Another test to determine the role of the apparatus alone would be to redo the experiments at zero pressure {\it inside} the DAC, to better differentiate
the roles of the actual pressure versus the presence of the DAC itself.

Ultimately, however, for a convincing demonstration of the utility of magnetization measurements on hydride materials in DACs, one
would need to follow the same protocol in a known superconductor as was used in the hydrides, and properly document the 
background subtractions that are required.

 \section{conclusion}
 In Ref.~\cite{tiny}, the authors conducted trapped flux experiments on a sample of a known superconducting material, attempting to mimic
 the conditions in the experiments for hydrides under pressure \cite{etrappedp}, by picking a tiny sample of volume comparable to the
 hydride samples, and subjecting it to pressure in a diamond anvil cell. 
 
 Contrary to the statements in Refs.~\cite{budko, canfield, talan,tiny} by Bud'ko and coauthors that our model is unphysical and incorrect and wrong and that we selectively
 hide/delete data, our model reproduces the observed behavior in this superconducting
 tiny sample remarkably well, as shown in  Fig.~\ref{figure2}. Upon picking the parameters in the model to fit the low field behavior of the ZFC moment,
 our model predicts the behavior of the small field FC moment as well as the overall behavior of the ZFC and FC moments including the values
 where the moments reach saturation remarkably well. The maximum discrepancy in the values of calculated moment vs measured 
 moment in Fig.~\ref{figure2} is $25\%$. This indicates that the model captures the essential physics of trapped moments in 
 type II superconductors even for tiny samples such as those used in the hydride experiments.
 
 In contrast, we showed that our model cannot describe the behavior of the moments reported for the hydride sample H$_3$S~\cite{etrappedp}  
 for any set of parameters in the model.
 Choosing $H^*=0.7T$ as was done in Ref.~\cite{etrappedp}, to fit approximately the fields where the moments reach
 saturation, gives rise to major discrepancies in the calculated versus measured values for lower fields, as shown in Fig.~\ref{figure3};
the discrepancy in the values of calculated moment vs measured 
 moment in Fig.~\ref{figure3}  is up to $800\%$, 30 times larger than for the superconducting sample in Fig.~\ref{figure2}. 
 If $H^*$ is instead chosen to approximately fit the low field behavior, there are major differences for the high field behavior,
 as shown in Figs.~\ref{figure4} and~\ref{figure5}. 
 Furthermore,  the quadratic behavior of ZFC moment with field at small field  that is seen in this
 tiny sample~\cite{tiny}, as well as all the superconductor samples studied by these authors earlier~\cite{canfield}, is perfectly fitted by our model. The 
 linear behavior seen in the hydrides that is not fitted by our model is in stark contrast with the behavior of all the superconducting samples, including the  sample studied in Ref. \cite{tiny} in the DAC under pressure 2.2 GPa, as we discussed in Sect. III. 
 
 Furthermore, Ref.~\cite{tiny} showed that when the tiny sample is measured in a DAC under pressure
 significant other issues come into play. The magnetization under an applied field as well as the noise show 
 large differences compared to the measurements without the DAC. To be relevant to the interpretation of hydride results,
 the protocols used, in particular background subtraction, should be the same. In particular, Ref.~\cite{tiny}  emphasized the need for point-by-point
 background subtraction, it is not clear what was done in Ref.~\cite{etrappedp}.

We conclude that
 the remnant moments measured in hydrides~\cite{etrappedp}, shown here in Fig.~\ref{figure1} lower panel, do not show
 behavior that is consistent with the behavior seen in the tiny superconducting sample studied in Ref. \cite{tiny}, neither at ambient pressure nor under 
 pressure. The most likely possibility is that the remnant moments measured in Ref. \cite{etrappedp}    originate in 
 other phenomena or experimental artifacts unrelated to superconductivity,  in accordance with the conclusions reached
 in our earlier papers on this subject~\cite{hmtrapped,hmtrapped2}.

  \begin{acknowledgments}
 FM was supported in part by the Natural Sciences and Engineering Research Council of Canada (NSERC) and by Alberta Innovates.
  \end{acknowledgments}

 \end{document}